\documentclass[%
 reprint,
 amsmath,amssymb,
 aps,
]{revtex4-2}

\usepackage{graphicx}
\usepackage{dcolumn}
\usepackage{bm}
\usepackage{algorithm}
\usepackage{algorithmic}
\RequirePackage[colorlinks=true,linkcolor=blue,urlcolor=blue,citecolor=blue]{hyperref}

\begin{document}

\preprint{APS/123-QED}

\title{Active Learning with Variational Quantum Circuits for Quantum Process Tomography}




\author{Jiaqi Yang}%
\email{yangjiaqi@mail.ustc.edu.cn}
\affiliation{University of Science and Technology of China, Hefei 230027, China }
\author{Xiaohua Xu}%
\email{xiaohuaxu@ustc.edu.cn}
\thanks{Corresponding author}
\affiliation{University of Science and Technology of China, Hefei 230027, China }
\author{Wei Xie}%
\email{xxieww@ustc.edu.cn}
\thanks{Corresponding author}
\affiliation{University of Science and Technology of China, Hefei 230027, China }

\date{\today}

\begin{abstract}
Quantum process tomography (QPT) is a fundamental tool for fully characterizing quantum systems. It relies on querying a set of quantum states as input to the quantum process. Previous QPT methods typically employ a straightforward strategy for randomly selecting quantum states, overlooking differences in informativeness among them. In this work, we propose a general active learning (AL) framework that adaptively selects the most informative subset of quantum states for reconstruction. We design and evaluate various AL algorithms and provide practical guidelines for selecting suitable methods in different scenarios. In particular, we introduce a learning framework that leverages the widely-used variational quantum circuits (VQCs) to perform the QPT task and integrate our AL algorithms into the query step. We demonstrate our algorithms by reconstructing the unitary quantum processes resulting from random quantum circuits with up to seven qubits. Numerical results show that our AL algorithms achieve significantly improved reconstruction, and the improvement increases with the size of the underlying quantum system. Our work opens new avenues for further advancing existing QPT methods.
\end{abstract}
\maketitle


\section{Introduction}

Quantum process tompography (QPT), used for reconstruction of an unknown quantum process from measurement data, is a fundamental tool for the diagnostic and full characterization of quantum systems \cite{chuang1997prescription}. Standard quantum process tomography (SQPT) relies on an informationally complete set of quantum states as input, which inevitably leads to an algorithmic complexity in terms of the number of measurements and classical post-processing that scales exponentially with the system size \cite{d2001quantum}.  Due to these limitations, SQPT has only been experimentally implemented up to three qubits \cite{krinner2020benchmarking,munoz2020single}. However, with the rapid advancement of quantum computing hardware, there is a growing demand for scalable QPT schemes \cite{guo2019general,wu2021strong}.

Inspired by machine learning techniques, various methods have been proposed to alleviate the exponential scaling problem. Some statistical methods such as maximum-likelihood estimation \cite{smolin2012efficient}, Bayesian estimation \cite{granade2016practical}, and projected-least-squares \cite{surawy2022projected} techniques have been adapted for QPT. A major challenge in current applications is the experimental and computational cost of estimating large-scale systems. A class of methods partially address this challenge, including compressed-sensing QPT \cite{flammia2012quantum,rodionov2014compressed}, tensor network QPT \cite{torlai2023quantum}, but at the cost of assuming specific structures of the unknown quantum process. In recent years, quantum machine learning combined with variational quantum circuits (VQCs) has received extensive attention \cite{torlai2018neural,abbas2021power}. Another promising class of methods focuses on reconstructing a unitary quantum process by inverting the dynamics using variational algorithms \cite{carolan2020variational,xue2022variational}. 

Indeed, although these QPT methods employ different techniques for reconstruction, they all involve a crucial step: \textit{query}, which requires selecting a subset of quantum states from the informationally complete set as inputs to the unknown process. It is worth noting that quantum states with greater informativeness typically improving reconstruction more efficiently. However, previous works typically use a straightforward strategy to select the subset randomly, overlooking differences in informativeness among quantum states. Since querying the quantum system requires multiple experiments that can be prohibitively costly, it is always the case that there are not enough quantum states for high-quality reconstruction, especially in large-scale quantum systems.

Building on this observation, this work does not aim to replace existing QPT methods, but rather proposes a general framework that further enhances them by optimizing the crucial query step. Specifically, we focus on the informativeness of quantum states, aiming to select a subset of quantum states that improves the reconstruction most efficiently. Active learning (AL) is a promising machine learning paradigm that helps reduce the annotation costs by intelligently selecting a subset of informative samples for labeling \cite{settles2009active}. With the aim of reducing experiment costs, AL has been extended from computer science to physics and applied successfully in various subfields, such as condensed matter physics \cite{zhang2019active,yao2020active}, high-energy physics \cite{caron2019constraining}, and quantum information \cite{ding2020retrieving,lange2023adaptive,dutt2023active}. 

In this paper, we propose a general AL framework that adaptively selects the most informative subset of quantum states for reconstruction. To the best of our knowledge, this is the first framework that explicitly utilizes the differences in informativeness among quantum states for QPT. We design and evaluate three various types of AL algorithms with the aim of offering choices suited to different scenarios, including committee-based, uncertainty-based, and diversity-based, each exhibiting distinct advantages in terms of performance and computational cost \cite{kumar2020active}. In particular, we introduce a learning framework that leverages the widely-used VQCs to perform the QPT task and integrate our AL algorithms into the query step. Notably, we emphasize our AL framework is general as it can in principle be applied in the query step of any QPT method. We demonstrate our algorithms by reconstructing the unitary quantum processes resulting from random quantum circuits with up to seven qubits. Numerical results show that our AL algorithms achieve significantly improved reconstruction compared to the baseline method, which randomly selects a subset of quantum states. Moreover, the improvement increases with the size of the underlying quantum system.

This paper is organized as follows. In Sec.~\ref{sec:2}, we introduce how widely-used VQCs are leveraged to perform the QPT task. We then provide a detailed explanation of the learning framework. In Sec.~\ref{sec:3}, we propose a general AL framework that extends the learning framework by optimizing its query step through the exploitation of differences in informativeness among quantum states. We further design various AL algorithms. In Sec.~\ref{sec:4}, we demonstrate our algorithms with numerical simulations on the quantum processes of randomly generated circuits with up to seven qubits. We conclude in Sec.~\ref{sec:5}.

\section{Learning Unitary Quantum Process with Variational Quantum Circuits}\label{sec:2}

\begin{figure*}[!t]
	\centering
	\includegraphics[width=7.1in]{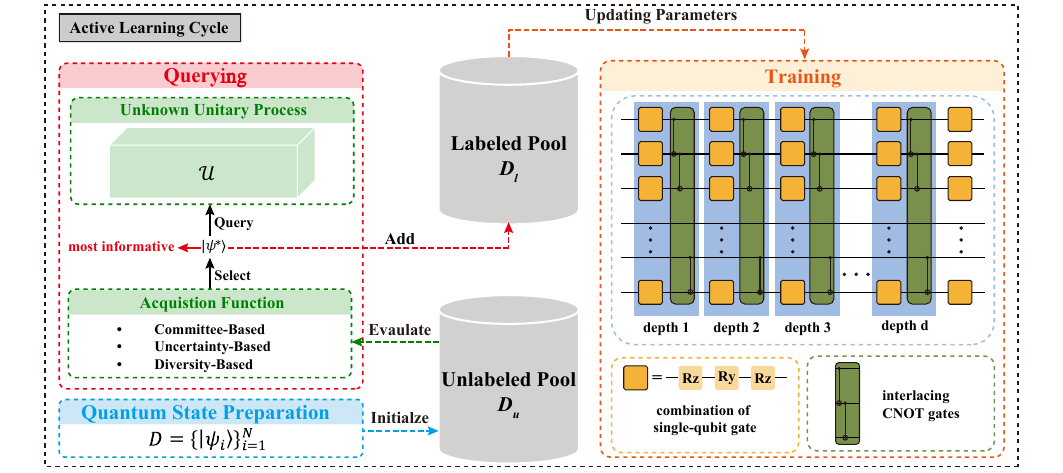}
	\caption{Schematic of our proposed active learning framework for quantum process tomography. The framework consists of three key components: \textit{Quantum State Preparation}, \textit{Querying}, and \textit{Training}. In the \textit{Quantum State Preparation} step, we prepare an informationally complete set of quantum states to initialize the unlabeled pool \(D_u\). We show our active learning cycle as follows: it consists of two steps, i.e., \textit{Querying} and \textit{Training}. In the \textit{Querying} step, we select the most informative quantum state \( |\psi^\star\rangle\) via the designed \textit{acquisition function} (including various types, e.g., Committee-Based, Uncertainty-Based, and Diversity-Based). We then query the unknown unitary process \(\mathcal{U}\) to obtain its true label \( \mathcal{U}|\psi^\star\rangle\). The newly labeled quantum state \((|\psi^\star\rangle,\mathcal{U}|\psi^\star\rangle)\) is subsequently added to the labeled pool \(D_l\). In the \textit{Training} step, we retrain the VQC model using the updated labeled pool \(D_l\) to better approximate the unknown unitary process \(\mathcal{U}\). This cycle repeats until the labeling budget is exhausted or the VQC model achieves satisfactory performance.} 
    \label{fig:AL}
\end{figure*}

\subsection{Model}
In this section, we discuss how to leverage widely-used VQCs to perform the QPT task and provide a detailed explanation of the learning framework. 

\textit{Unknown Quantum Process.---}  Let $\mathcal{U}$ be an unknown unitary quantum process. It is reasonable to compile it into a VQC \cite{khatri2019quantum,xue2022variational}. More specifically, we can approximate \(\mathcal{U}\) using a VQC denoted by \(\mathcal{C}(\boldsymbol{\theta})\), where \(\boldsymbol{\theta}\) represents the parameters to be optimized. Consequently, all the information about \( \mathcal{U} \) is encoded in the parameters \( \boldsymbol{\theta} \), enabling us to systematically reconstruct \( \mathcal{U} \) from these parameters using a classical computer.

\textit{Variational Quantum Circuit Ansatz.---} The ansatz of VQC is shown in Fig.~\ref{fig:AL}. It is composed of interlaced layers using single-qubit Pauli rotation gates and two-qubit CNOT gates. This structure is designed to quickly generate entanglement between qubits, thus making it possible to approximate complicated quantum processes. Each two-qubit layer is counted as a depth and each single-qubit layer contains three rotational gates (\(R_z\), \(R_y\), and \(R_z\)) on each qubit, where
\[
\renewcommand{\arraystretch}{1.3} 
R_y(\theta) = 
\begin{pmatrix}
\cos\frac{\theta}{2} & -\sin\frac{\theta}{2} \\
\sin\frac{\theta}{2} & \cos\frac{\theta}{2}
\end{pmatrix}, \quad
R_z(\theta) = 
\begin{pmatrix}
\textnormal{e}^{-\textnormal{i}\frac{\theta}{2}} & 0 \\
0 & \textnormal{e}^{\textnormal{i}\frac{\theta}{2}}
\end{pmatrix}.
\renewcommand{\arraystretch}{1} 
\]
The combination of \(R_z\), \(R_y\), and \(R_z\) rotations can realize any single-qubit rotation by adjusting the parameters. For such an $n$-qubit circuit with depth \( k \), the total number of parameters is \( 3n(k + 1) \). 

\subsection{Learning Framework}
The learning framework encompasses three key components: quantum state preparation, querying, and training, as shown in Fig.~\ref{fig:AL}.

\textit{Quantum State Preparation.---} A quantum state is prepared by applying a unitary transformation \( U \) to the initial state: \( |0\rangle^{\otimes n} \rightarrow U|0\rangle^{\otimes n} \). Specifically, the unitary transformation \( U \) is selected from tensor products of single-qubit Pauli operators \( \left\{ I, X, Y, Z \right\}^{\otimes n} \). We generate an informationally complete set of quantum states, denoted as \( D = \left\{ \left| \psi_1 \right\rangle, \left| \psi_2 \right\rangle, \ldots, \left| \psi_N \right\rangle \right\} \), where \( N = 4^n \).

\textit{Querying.---} We refer to the unknown unitary process \( \mathcal{U} \) as an oracle that can be queried. When a quantum state \( |\psi\rangle \) is queried to the oracle, it returns the state \(|\psi^{\mathrm{ideal}}\rangle=\mathcal{U}|\psi\rangle \) as its true label. This procedure allows us to construct a training dataset.

\textit{Training.---} Given a training dataset \( D_l = \left\{ \left( \left| \psi_1 \right\rangle, \left| \psi_1^{\mathrm{ideal}} \right\rangle \right), \left( \left| \psi_2 \right\rangle, \left| \psi_2^{\mathrm{ideal}} \right\rangle \right), \ldots, \left( \left| \psi_m \right\rangle, \left| \psi_m^{\mathrm{ideal}} \right\rangle \right) \right\} \), the loss function is defined as the mean square error:
\begin{align} \label{eq:1}
    \mathcal{L}(\boldsymbol{\theta}) &= \frac{1}{m} \sum_{i=1}^m \left\| \left|\psi_i(\boldsymbol{\theta})\right\rangle - \left|\psi_i^{\mathrm{ideal}}\right\rangle \right\|^2 \notag \\
    &= \frac{1}{m} \sum_{i=1}^m \bigg( \left\langle \psi_i(\boldsymbol{\theta}) \mid \psi_i(\boldsymbol{\theta}) \right\rangle + \left\langle \psi_i^{\mathrm{ideal}} \mid \psi_i^{\mathrm{ideal}} \right\rangle \notag \\
    &\quad - \left\langle \psi_i(\boldsymbol{\theta}) \mid \psi_i^{\mathrm{ideal}} \right\rangle - \left\langle \psi_i^{\mathrm{ideal}} \mid \psi_i(\boldsymbol{\theta}) \right\rangle \bigg) \notag \\
    &= \frac{2}{m} \sum_{i=1}^m \left[ 1 - \mathrm{Re} \left( \left\langle \psi_i^{\mathrm{ideal}} \mid \psi_i(\boldsymbol{\theta}) \right\rangle \right) \right],
\end{align}
where \( \left|\psi_i\left(\boldsymbol{\theta}\right)\right\rangle = \mathcal{C}\left(\boldsymbol{\theta}\right)\left|\psi_i\right\rangle \) denotes the predicted label, and \( \|\cdot\|\) denotes the Euclidean norm. The inner product on the right-hand side can be efficiently estimated using a generalized SWAP-test method \cite{xue2022variational}.

To minimize the loss function, we employ a gradient-based optimization method to iteratively update the parameters \(\boldsymbol{\theta}\). Using the chain rule, the gradient is computed as
\begin{equation} \label{eq:2}
    \frac{\partial \mathcal{L}(\boldsymbol{\theta})}{\partial \theta_i} = \frac{\partial \mathcal{L}(\boldsymbol{\theta})}{\partial \mathcal{O}(\boldsymbol{\theta})} \frac{\partial \mathcal{O}(\boldsymbol{\theta})}{\partial \theta_i},
\end{equation}
where \(\mathcal{O}\) represents the observable corresponding to the generalized SWAP test. The first term, \(\frac{\partial \mathcal{L}(\boldsymbol{\theta})}{\partial \mathcal{O}(\boldsymbol{\theta})}\), can be computed classically once the value of \(\mathcal{O}(\boldsymbol{\theta})\) is obtained from the quantum computer. The second term, \(\frac{\partial \mathcal{O}(\boldsymbol{\theta})}{\partial \theta_i}\), is computed on the quantum computer using the parameter-shift rule \cite{mitarai2018quantum}:
\begin{equation} \label{eq:3}
    \frac{\partial \mathcal{O}(\boldsymbol{\theta})}{\partial \theta_i} = \frac{1}{2} \left[ \mathcal{O}(\boldsymbol{\theta}_i^+) - \mathcal{O}(\boldsymbol{\theta}_i^-) \right],
\end{equation}
where \(\boldsymbol{\theta}_i^{\pm}\) denotes the parameter vector \(\boldsymbol{\theta}\) with the \(i\)-th parameter shifted by \(\pm \frac{\pi}{2}\).

\section{Method}\label{sec:3}
In this section, we first analyze the inefficiencies in the crucial query step. To address this issue, we then propose a general AL framework that explicitly utilizes the differences in informativeness among quantum states to optimize the query step. Finally, we design various AL algorithms with the aim of offering choices suited to different scenarios. Notably, we emphasize our AL framework is general and can in principle be applied in the query step of any QPT method, not just limited to the VQCs considered here.

\subsection{Active Learning} 
Considering that querying a quantum system is prohibitively costly since it requires multiple experiments, our goal is to learn the parameters \( \boldsymbol{\theta} \) with error \( \epsilon \) from training dataset \( \{(|\psi_i\rangle, |\psi_i^{\mathrm{ideal}}\rangle)\}_{i=1}^m \), while minimizing the number of labeled quantum states \( m \)  (i.e., queries to the unknown process \(\mathcal{U}\)). However, in the query step, it is common to select a subset of quantum states randomly. This straightforward strategy generally requires a large number of labeled quantum states to train a VQC to achieve satisfactory performance. As the system scale increases, this challenge becomes more pronounced---it is always the case that there are insufficient labeled quantum states to train a VQC with desirable performance for high-quality reconstruction. In fact, each quantum state contains different informativeness, with those having greater informativeness typically improving training more efficiently. This random selection strategy does not take this into account, resulting in training inefficiencies. To overcome this limitation, we focus on the informativeness of quantum states, aiming to intelligently select a subset of quantum states that improves the training most efficiently.   

\textit{Acquisition Function.---} AL is a machine learning paradigm that promises to reduce annotation costs by intelligently selecting a subset of informative samples for labeling \cite{settles2009active}. The selection of such a subset relies on a so-called \textit{acquisition function}. This function plays a central role in AL. It operates by evaluating samples based on criteria such as committee disagreement, prediction uncertainty, or diversity, which essentially estimate the informativeness of samples from different perspectives \cite{kumar2020active}. Specifically, the acquisition function evaluates the informativeness of each unlabeled sample and ranks them accordingly. The highest-ranking sample is then selected for labeling and subsequently added to the training dataset for updating the model. This process is repeated, gradually increasing the number of labeled samples and improving the model’s performance. 

\textit{Active Learning Framework.---} A schematic overview of our AL framework is provided in Fig.~\ref{fig:AL}. It consists of the following steps:

\begin{enumerate}
    \item \textit{Initialize}: Begin by preparing an unlabeled dataset \(D_u\) that consists of an informationally complete set of quantum states (i.e., a set of quantum states sufficient to fully characterize the quantum process). Also, initialize an empty labeled pool \(D_l\). It will be gradually expanded as labeled quantum states are added during the AL process. Finally, set up a base VQC model \(\mathcal{C}(\boldsymbol{\theta})\), which will be iteratively trained to learn the unknown quantum process \( \mathcal{U} \).
    \item \textit{Evaluate}: Use the acquisition function to evaluate the informativeness of each quantum state in the unlabeled pool \(D_u\). Then, rank all the states according to these information values, thereby effectively identifying the most informative quantum state for labeling.
    \item \textit{Select}: Select the highest-ranking quantum state \(|\psi^\star\rangle\) from the unlabeled pool \(D_u\). This carefully chosen quantum state is expected to contribute most efficiently to improving the VQC model’s performance.
    \item \textit{Query}: Query the unknown unitary process \( \mathcal{U} \) by inputting the selected quantum state \(|\psi^\star\rangle\), and obtain the corresponding output \(\mathcal{U}|\psi^\star\rangle \), which serves as its true label.
    \item \textit{Add}: Add the newly labeled quantum state \((|\psi^\star\rangle, \mathcal{U}|\psi^\star\rangle)\) to the labeled pool \(D_l\). This step enables the labeled pool \(D_l\) to gradually expand throughout the AL process, thereby improving the training of the VQC model \(\mathcal{C}(\boldsymbol{\theta})\) to better approximate the unknown quantum process \(\mathcal{U}\).
    \item \textit{Update}: Retrain the VQC model \(\mathcal{C}(\boldsymbol{\theta})\) using the updated labeled pool \(D_l\), which now includes the newly added quantum state. 
\end{enumerate}

The process follows an iterative procedure in which quantum states are sequentially labeled to improve model performance. At each AL step, the most informative quantum state is selected based on the acquisition function. This AL cycle of \textit{Evaluate}---\textit{Select}---\textit{Query}---\textit{Add}---\textit{Update} repeats until the labeling budget is exhausted or a satisfactory model performance is achieved. 

Based on various acquisition functions, current AL methods can be broadly classified into committee-based, uncertainty-based and diversity-based \cite{kumar2020active}.  As mentioned earlier, these approaches essentially estimate the informativeness of samples from different perspectives. With the aim of offering choices suited to different scenarios, we design various AL algorithms to each category as follows.

\subsection{Query-by-Committee}
Query-by-committee (QBC) is a widely used AL approach \cite{freund1997selective,burbidge2007active}. Its basic idea is to build a committee of models from existing labeled data, and then selects the sample for which the committee disagree most to label. This is because the degree of disagreement among the committee members can serve as an estimate of the information value.

We construct a committee of \( n_\mathrm{VQC} \) models \( \boldsymbol{\Theta} = \{\boldsymbol{\theta}_1, \boldsymbol{\theta}_2, \ldots, \boldsymbol{\theta}_{n_\mathrm{VQC}}\} \), which are initialized with different parameters and trained on \( D_l \). In each AL step, each member of the committee \( \boldsymbol{\Theta} \) casts its vote based on their learned quantum process representations \( \{ \mathcal{C}(\boldsymbol{\theta}_1), \mathcal{C}(\boldsymbol{\theta}_2), \ldots, \mathcal{C}(\boldsymbol{\theta}_{n_\mathrm{VQC}}) \} \). The committee select the quantum state for which all models disagree most on the predictions 
\( \{|\psi(\boldsymbol{\theta}_1)\rangle,|\psi(\boldsymbol{\theta}_2)\rangle\}, \ldots, |\psi(\boldsymbol{\theta}_{n_\mathrm{VQC}})\rangle\} \). A common strategy to calculate the level of disagreement is the variance. More precisely, for each unlabeled quantum state \( |\psi\rangle\in D_u \), we compute the variance of the \( n_\mathrm{VQC} \) individual predictions and then select the quantum state with the maximum variance, that is, 
\begin{equation} \label{eq:4}
    |\psi^\star\rangle = \underset{|\psi\rangle\in D_u}{\operatorname{argmax}} \frac{1}{n_{\mathrm{VQC}}} \sum_{k=1}^{n_{\mathrm{VQC}}} d\left(|\psi(\boldsymbol{\theta}_k)\rangle, |\psi(\boldsymbol{\Theta})\rangle\right),
\end{equation}
where \( |\psi({\boldsymbol{\Theta}})\rangle\propto\sum_{k=1}^{n_{\mathrm{VQC}}} |\psi({\boldsymbol{\theta}_k})\rangle \), and \( d \) represents the Euclidean distance. The pseudo-code for QBC is shown in Algorithm \ref{alg:QBC}.

\begin{algorithm}[H]
\caption{Query-by-Committee (QBC)}
\label{alg:QBC}
\textbf{Input}: Labeled quantum states set \(D_l\), unlabeled quantum states set \(D_u\), number of committee members \(n_\mathrm{VQC}\)\\
\textbf{Output}: Selected quantum state \( |\psi^\star\rangle \) with the maximum variance
\begin{algorithmic}[1] 
\STATE Construct a committee of \( n_\mathrm{VQC} \) models \( \boldsymbol{\Theta} = \{\boldsymbol{\theta}_1, \boldsymbol{\theta}_2, \ldots, \boldsymbol{\theta}_{n_\mathrm{VQC}}\} \), trained on \(D_l\)
\FOR{\( |\psi\rangle \in D_u \)}
    \FOR{\( k = 1 \) to \( n_\mathrm{VQC} \)}
    \STATE \( |\psi(\boldsymbol{\theta}_k)\rangle \leftarrow \mathcal{C}(\boldsymbol{\theta}_k)|\psi\rangle \)
    \ENDFOR
    \STATE Calculate the variance of \( \{|\psi(\boldsymbol{\theta}_k)\rangle\}_{k=1}^{n_\mathrm{VQC}} \) using Eq.~(\ref{eq:4})
\ENDFOR
\end{algorithmic}
\end{algorithm}

\subsection{Expected Model Change Maximization}
Expected model change maximization (EMCM) is also a popular AL approach that aims to select the sample resulting in the largest change to the current model \cite{cai2013maximizing}. The underlying rationale is that samples causing significant changes to the model are highly informative. They help reveal regions of the data distribution that the current model has not yet fully captured. For simplicity, the model change is typically measured as the difference between the current model parameters and the updated parameters after training with the enlarged training dataset. Inspired by Stochastic Gradient Descent (SGD) update rule, the change is estimated as the gradient of the loss with respect to an unlabeled sample.

Specifically, we employ SGD to update the parameters \( \boldsymbol{\theta} \). The parameters \( \boldsymbol{\theta} \) are updated iteratively according to the negative gradient descent of the loss \(\ell(\boldsymbol{\theta})\) with respect to each training data \( (|\psi_i\rangle, |\psi_i^{\mathrm{ideal}}\rangle) \in D_l \):
\begin{equation}
    \boldsymbol{\theta} := \boldsymbol{\theta} - \alpha \nabla_{\boldsymbol{\theta}} \ell_{|\psi_i\rangle}(\boldsymbol{\theta}) \quad \text{for } i = 1, 2, \ldots, m,
     \label{eq:5}
\end{equation}
where \(\alpha\) denotes the learning rate.

Here, we consider the update rule in AL process. Assume an unlabeled quantum state \( |\psi\rangle\in D_u \), which is then labeled as \( |\psi^\mathrm{ideal}\rangle\) and added to training dataset.
The loss function on the enlarged training dataset \( D_l^+ = D_l\cup(|\psi\rangle,|\psi^\mathrm{ideal}\rangle) \) then becomes:
\begin{small}
\begin{equation}
    \mathcal{L}_{D_l^+} = \frac{1}{|D_l^+|} \left[ \sum_{|\psi_i\rangle \in D_l} \left\| |\psi_i\rangle - |\psi_i^\mathrm{ideal}\rangle \right\|^2 + \underbrace{\left\| |\psi\rangle - |\psi^\mathrm{ideal}\rangle \right\|^2}_{=:\ell_{|\psi\rangle}(\boldsymbol{\theta})} \right].
    \label{eq:6}
\end{equation}
\end{small}
Thus, the model obtained by minimizing the loss function is changed due to the inclusion of the new training data \((|\psi\rangle, |\psi^\mathrm{ideal}\rangle)\). We measure the model change as the parameter change using the gradient of the loss at \( |\psi\rangle \). The derivative of the loss \( \ell_{|\psi\rangle}(\boldsymbol{\theta}) \) with respect to the parameters \( \theta_i \) at \( |\psi\rangle \) is formulated as:
\begin{equation} \label{eq:7}
    \begin{aligned}
    \frac{\partial \ell_{|\psi\rangle}(\boldsymbol{\theta})}{\partial \theta_i} 
    &= \frac{\partial \ell_{|\psi\rangle}(\boldsymbol{\theta})}{\partial \mathcal{O}(\boldsymbol{\theta})} \frac{\partial \mathcal{O}(\boldsymbol{\theta})}{\partial \theta_i} \nonumber \\
    &= \frac{\partial \ell_{|\psi\rangle}(\boldsymbol{\theta})}{\partial \mathcal{O}(\boldsymbol{\theta})} \left[ \frac{1}{2} \left( \mathcal{O}(\boldsymbol{\theta}_i^+) - \mathcal{O}(\boldsymbol{\theta}_i^-) \right) \right],
    \end{aligned}
\end{equation}
where \( \mathcal{O}(\boldsymbol{\theta}) = \mathrm{Re} \left( \left\langle \psi^{\mathrm{ideal}} \mid \psi(\boldsymbol{\theta}) \right\rangle \right) \). Since the true label \(|\psi^\mathrm{ideal}\rangle\) is actually unknown before querying in practice, we construct an ensemble \(\boldsymbol{\Theta} = \{\boldsymbol{\theta}_1, \boldsymbol{\theta}_2, \ldots, \boldsymbol{\theta}_{n_\mathrm{VQC}}\}\) to estimate the prediction \(\{|\psi(\boldsymbol{\theta}_1)\rangle, |\psi(\boldsymbol{\theta}_2)\rangle, \ldots, |\psi(\boldsymbol{\theta}_{n_\mathrm{VQC}})\rangle\}\), and use the expected model change to approximate the true model change, that is:
\begin{align} \label{eq:8}
    \frac{\partial \ell_{|\psi\rangle}(\boldsymbol{\theta})}{\partial \theta_i} 
    &= \frac{1}{n_\mathrm{VQC}} \sum_{k=1}^{n_\mathrm{VQC}} \frac{\partial \ell_{|\psi\rangle}(\boldsymbol{\theta})}{\partial \hat{\mathcal{O}}_k(\boldsymbol{\theta})} \frac{\partial \hat{\mathcal{O}}_k(\boldsymbol{\theta})}{\partial \theta_i} \nonumber \\
    &= \frac{1}{2n_\mathrm{VQC}} \sum_{k=1}^{n_\mathrm{VQC}} \frac{\partial \ell_{|\psi\rangle}(\boldsymbol{\theta})}{\partial \hat{\mathcal{O}}_k(\boldsymbol{\theta})} \left[ \hat{\mathcal{O}}_k(\boldsymbol{\theta}_i^+) - \hat{\mathcal{O}}_k(\boldsymbol{\theta}_i^-) \right],
\end{align}
where \( \hat{\mathcal{O}}_k(\boldsymbol{\theta}) = \mathrm{Re} \left( \left\langle \psi(\boldsymbol{\theta}_k) \mid \psi(\boldsymbol{\theta}) \right\rangle \right)\) represents an estimate of \(\mathcal{O}\).

The goal of the sampling strategy is to select the quantum state that could maximally change the current model. Therefore, the acquisition function can be formulated as:
\begin{equation} \label{eq:9}
    |\psi^\star\rangle = \underset{|\psi\rangle\in D_u}{\operatorname{argmax}} \left\| \nabla_{\boldsymbol{\theta}} \ell_{|\psi\rangle}(\boldsymbol{\theta}) \right\|.
\end{equation}
The pseudo-code for EMCM is shown in Algorithm~\ref{alg:EMCM}.

\begin{algorithm}[H]
\caption{Expected Model Change Maximization (EMCM)}
\label{alg:EMCM}
\textbf{Input}: Labeled quantum states set \( D_l\), unlabeled quantum states set \( D_u \), base model \( \mathcal{C}(\boldsymbol{\theta}) \) trained on \( D_l \), number of ensemble models \(n_\mathrm{VQC}\)\\
\textbf{Output}: Selected quantum state \( |\psi^\star\rangle \) with the maximum gradient norm
\begin{algorithmic}[1] 
\STATE Construct an ensemble of \( n_\mathrm{VQC} \) models \( \boldsymbol{\Theta} = \{\boldsymbol{\theta}_1, \boldsymbol{\theta}_2, \ldots, \boldsymbol{\theta}_{n_\mathrm{VQC}}\} \), trained on \( D_l \)
\FOR{\( |\psi\rangle \in D_u \)}
    \STATE \(|\psi(\boldsymbol{\theta})\rangle\leftarrow\mathcal{C}(\boldsymbol{\theta})|\psi\rangle \)
    \FOR{\( k = 1 \) to \( n_\mathrm{VQC} \)}
    \STATE \( |\psi(\boldsymbol{\theta}_k)\rangle \leftarrow \mathcal{C}(\boldsymbol{\theta}_k)|\psi\rangle \)
    \ENDFOR
    \STATE Calculate the gradient norm \( \left\| \nabla_{\boldsymbol{\theta}} \ell_{|\psi\rangle}(\boldsymbol{\theta}) \right\| \) using Eq.(\ref{eq:8})
\ENDFOR
\end{algorithmic}
\end{algorithm}

\subsection{Greedy Sampling}
Instead of finding the most informative sample based on the learned model, as in QBC and EMCM, greedy sampling (GS) is a model-free approach that selects the sample based on its geometric characteristics in the feature space \cite{yu2010passive}. Its basic idea is to greedily select an unlabeled sample that is located far from the previously labeled samples, thereby ensuring that the selected samples are scattered across the entire input space rather than concentrating in a small local region. This strategy effectively maximizes coverage and diversity of the selected samples, which are more likely to capture varied and complementary information. Clearly, GS allows for selecting a set of samples at the beginning of the AL process, making it easy to implement by avoiding the need to evaluate each unlabeled sample and update the model at each AL step.  

For each unlabeled quantum state \( |\psi\rangle \in D_u \), we compute its distance to all labeled quantum states in \( D_l \). The quantum state that maximizes the its distance to the labeled quantum state set is selected. Formally, 
\begin{equation} \label{eq:10}
    |\psi^\star\rangle = \underset{|\psi\rangle \in D_u}{\operatorname{argmax}} \underset{|\psi_i\rangle \in D_l}{\operatorname{min}} d(|\psi\rangle, |\psi_i\rangle),
\end{equation}
where \( d \) represents the Euclidean distance. The pseudo-code for GS is shown in Algorithm~\ref{alg:GS}.

\begin{algorithm}[H]
\caption{Greedy Sampling (GS)}
\label{alg:GS}
\textbf{Input}: Labeled quantum states set \( D_l \), unlabeled quantum states set \( D_u \) \\
\textbf{Output}: Selected the quantum state \( |\psi^\star\rangle \) with the maximum of the minimum distance
\begin{algorithmic}[1] 
\FOR{\(|\psi\rangle \in D_u \)}
    \FOR{\( |\psi_i\rangle \in D_l \)}
    \STATE Calculate the distance \(d(|\psi\rangle, |\psi_i\rangle)\)
    \ENDFOR
    \STATE Find the minimum distance \(\underset{|\psi_i\rangle \in D_l}{\operatorname{min}} d(|\psi\rangle, |\psi_i\rangle)\)
\ENDFOR
\end{algorithmic}
\end{algorithm}

\section{Numerical Experiments}\label{sec:4}
In this section, we validate our algorithms through numerical simulations on quantum processes generated by random quantum circuits. Notably, QPT has been theoretically proven to be a fundamentally difficult task \cite{mohseni2008quantum}. As a result, most state-of-the-art studies typically perform numerical simulations on systems with maximum sizes ranging from four and seven qubits \cite{surawy2022projected,ahmed2023gradient,levy2024classical}. Therefore, we conduct simulations with up to seven qubits to evaluate the scalability of our algorithms.

\subsection{Experimental Settings}

\begin{figure}[H]
    \centering
    \includegraphics[width=3.5in]{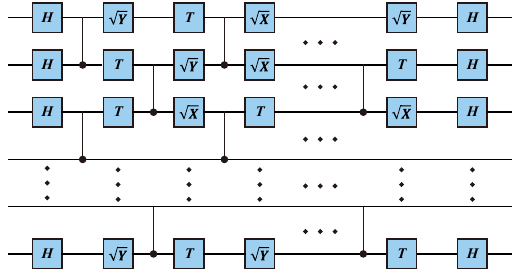} 
    \caption{Structure of a randomly generated quantum circuit.}
    \label{fig:RQC}
\end{figure}

\textit{Unknown Unitary Process.---} We consider the quantum process of randomly generated circuits, as shown in Fig.~\ref{fig:RQC}. The circuit begins and ends with Hadamard gates applied to each qubit. Each two-qubit layer is counted as a depth, including controllesd-phase (CZ) gates alternating between odd and even depths, and randomly chosen single-qubit gate (\(T\), \(\sqrt{X}\), or \(\sqrt{Y}\)). It is noted that such randomly generated quantum circuits are hard for efficient simulation on a classical computer \cite{boixo2018characterizing,bouland2019complexity}.

\textit{Baseline and Performance Metrics.---} We consider a baseline random (RAND) that selects a subset of quantum states randomly. Additionally, we use similarity between the target process \(\mathcal{U}\) and the approximate reconstruction \(\mathcal{C}(\boldsymbol{\theta})\) to measure performance, defined as
\begin{align}
    \textnormal{similarity}\left(\mathcal{U},\mathcal{C}(\boldsymbol{\theta})\right)=1-\frac{\|\mathcal{C}(\boldsymbol{\theta})-\mathcal{U}\|_F}{2\|\mathcal{U}\|_F}.
    \notag
\end{align}
Here, \(\|\cdot\|_F\) denotes the Frobenius norm of a matrix. The \(\text{similarity}(\mathcal{U},\mathcal{C}(\boldsymbol{\theta}))=1\) means \(\|\mathcal{C}(\theta)-\mathcal{U}\|_F=0\), in which case \(\mathcal{U}\)  can be perfectly reconstructed from \(\mathcal{C}(\boldsymbol{\theta})\).

To better evaluate our algorithms, we define an additional metric called \textit{improvement}, which is the ratio of the similarity achieved by our algorithms to that achieved by RAND, with the same number of labeled quantum states.

\textit{Other Experimental Parameters.---}  We conduct simulations with different numbers of qubits to evaluate the scalability of our algorithms. For an \(n\)-qubit quantum process (\(n\) ranges from 2 to 7), we set the VQC depth to 3, 5, 7, 8, 8, and 8 to learn the corresponding quantum processes. The number of committee members in QBC and the number of ensemble models in EMCM is empirically set to be 6. To avoid random fluctuation, each experiment is repeated 100 times and the averaged metrics are reported. 

\subsection{Experimental Results}
\begin{figure*}[!t]
	\centering
	\includegraphics[width=7.1in]{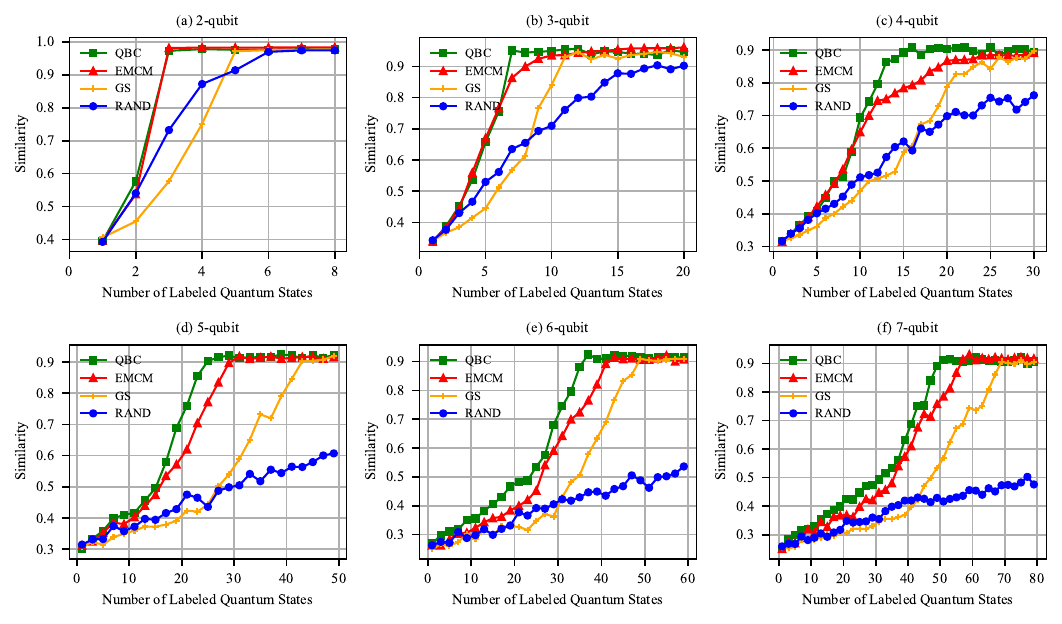}
	\caption{Similarity under different query strategies: QBC (green squares), EMCM (red triangles), GS (orange plus signs), and RAND ((blue circles). Subplots (a)–(f) correspond to 2- to 7-qubit systems, respectively.} 
	\label{fig:Similarity1}
\end{figure*}

\textit{Similarity Metric.---} Fig.~\ref{fig:Similarity1} shows the similarity comparison between our three proposed AL algorithms and the baseline RAND across varying numbers of labeled quantum states. The key observations are as follows.
\begin{itemize}
    \item Generally, as the number of labeled quantum states increased, all four algorithms achieved better performance. This is intuitive, as more training samples typically result in a more reliable model.
    \item QBC, EMCM, and GS converges faster than RAND, each achieving higher similarity with significantly fewer labeled quantum states. This clearly demonstrates that our algorithms are highly effective in selecting informative quantum states, thereby improving the reconstruction efficiency. 
    \item QBC and EMCM significantly outperformed RAND across all numbers of labeled quantum states. GS initially performs worse than RAND when only a few quantum states are labeled. However, as the number of labeled quantum states increases, GS eventually outperforms RAND. A possible explanation for the phenomena is as follows. At the initial steps, GS might select some outliers to maximize the distance from the previously labeled quantum states, which leads to a decrease in performance. Since the number of outliers is usually very small, their impact diminishes as more quantum states are labeled. It is worth noting that as the system size increases, reconstructing the quantum process requires a significantly larger number of labeled quantum states, which helps mitigate the impact of outliers.
\end{itemize}

\begin{figure}[H]
    \centering
    \includegraphics[width=3.5in]{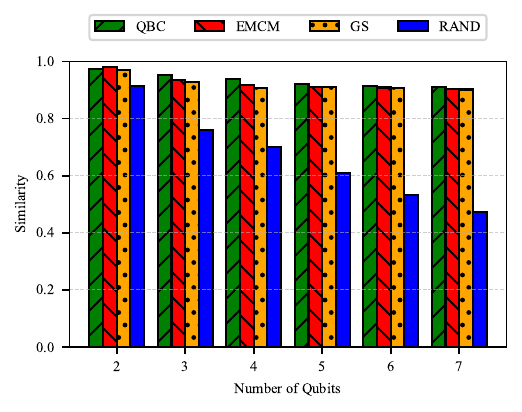} 
    \caption{Similarity under different query strategies: QBC (green with // hatch), EMCM (red with $\backslash$ hatch), GS (yellow with .. hatch), and RAND (blue).}
    \label{fig:Similarity2}
\end{figure}

Fig.~\ref{fig:Similarity2} shows the similarity comparison between our three proposed AL algorithms and the baseline RAND. Here, the number of labeled quantum states corresponds to the minimum required for all our AL algorithms to achieve a similarity of 0.9. Similarly, we conduct simulations on quantum processes with sizes ranging from 2 to 7 qubits. It is clearly from Fig.~\ref{fig:Similarity2} that as the system size increases, the performance gap between our AL algorithms and the baseline RAND becomes more significant. As the system size increases to 7 qubits, all AL algorithms achieve a similarity above 0.9. Meanwhile, the baseline RAND remains below 0.5.  

\begin{figure*}[!t]
	\centering
	\includegraphics[width=7.1in]{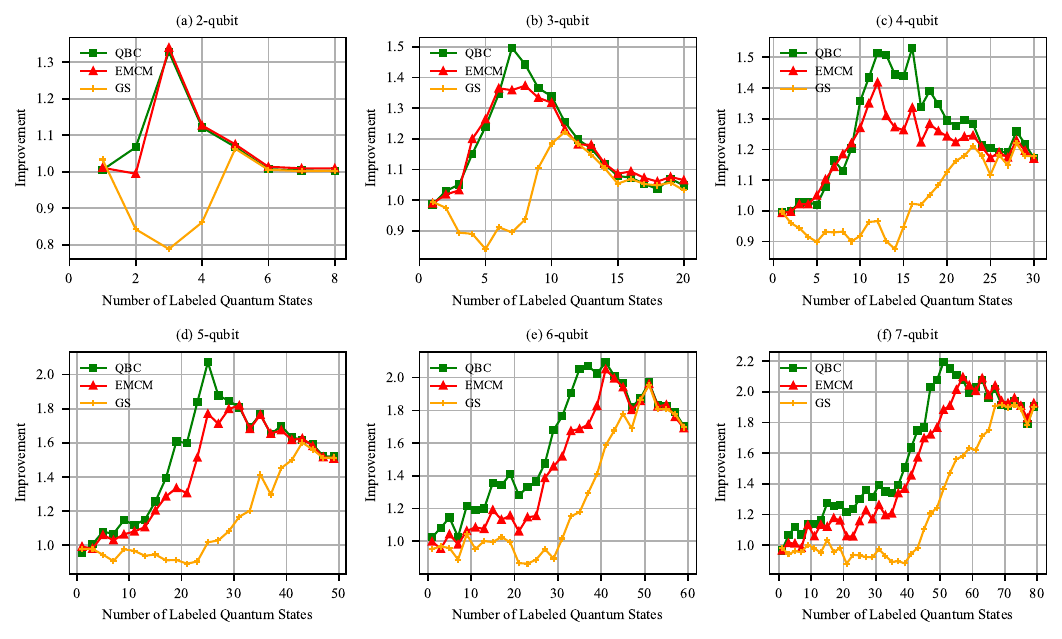}
	\caption{Improvement under different query strategies compared with the baseline RAND: QBC (green squares), EMCM (red triangles), and GS (orange plus signs). Subplots (a)–(f) correspond to 2- to 7-qubit systems, respectively.} 
	\label{fig:Improvement}
\end{figure*}

\textit{Improvement Metric.---} Fig.~\ref{fig:Improvement} shows the improvement comparison between our three proposed AL algorithms and the baseline RAND across varying numbers of labeled quantum states. A key finding is that as the number of labeled quantum states increases, the improvement of QBC and EMCM initially increases and then decreases. This is because, with sufficient labeled quantum states, even randomly selected quantum states can lead to well-trained models. In contrast, the improvement of GS initially decreases, which is consistent with our previous analysis, as GS may select outliers in the initial steps. As more quantum states are labeled, their impact diminishes and the improvement increases then decreases, similar to QBC and EMCM. 

\begin{figure}[!t]
    \centering
    \includegraphics[width=3.5in]{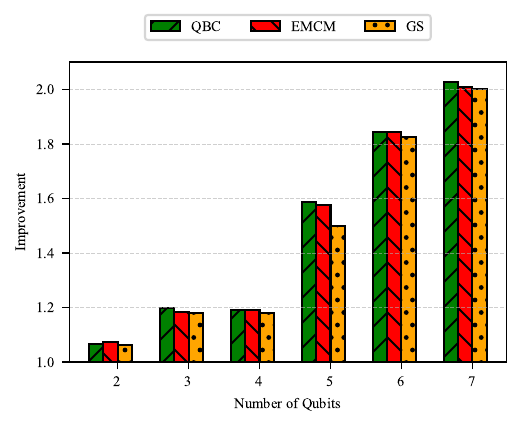} 
    \caption{Improvement under different query strategies: QBC (green with $\backslash$ hatch), EMCM (red with \ hatch), and GS (yellow with .. hatch).}
    \label{fig:Improvement2}
\end{figure}

Fig.~\ref{fig:Improvement2} shows the improvement of our three proposed AL algorithms compared to the baseline RAND. Here, the number of labeled quantum states corresponds to the minimum required for all our AL algorithms to achieve a similarity of 0.9. Similarly, we conduct simulations on quantum processes with sizes ranging from 2 to 7 qubits. It is obvious from Fig.~\ref{fig:Improvement2} that as the system size increases, the improvement of all AL algorithms also increases. This confirms that the improvement of our AL algorithms over the baseline RAND increases with the system size. As the system size increases to 7 qubits, the improvement of all AL algorithms exceeds 2. This indicates that, given the same number of labeled quantum states (i.e., queries to the unknown process), our AL algorithms can achieve more than twice the reconstruction similarity compared to the baseline RAND. These results further demonstrate the effectiveness of our AL algorithms, especially in large-scale quantum systems.


\textit{Practical Guidelines for Algorithms 
Selection.---} It is evident that QBC and EMCM achieved better performance than GS. However, both QBC and EMCM are model-based algorithms that require evaluating each unlabeled quantum states and updating the model at every AL step. Additionally, they rely on training multiple models, which is computationally expensive. In particular, QBC involves training a committee of models to quantify their disagreement in predictions, and EMCM involves training an ensemble of models to estimate the expected model change. In contrast to QBC and EMCM, GS is a model-free approach that selects quantum states based solely on their geometric characteristics. GS allows selecting a set of quantum states at the beginning of the AL process. This makes it easy to implement and computationally inexpensive since it avoids evaluating each unlabeled quantum state and updating the model at every AL step.

Considering performance, query cost, and computational cost, we can conclude that: In scenarios where query cost is the primary bottleneck (e.g., limited access to quantum hardware or high experimental costs), QBC and EMCM are generally preferable as they offer reliable performance even under a constrained query budget. Conversely, if computational cost is a major concern  (e.g., limited computational resources or large-scale simulations), GS is a more suitable choice as it can achieve reasonable performance with considerably less computational cost. 

\section{Conclusion} \label{sec:5}
In this work, we apply active learning to address the challenging problem of quantum process tomography. Rather than replacing existing QPT methods, our goal is to further enhance them by optimizing the crucial query step. We propose a general AL framework that adaptively selects a subset of informative quantum states to improve the reconstruction most efficiently. We design and evaluate three different types of AL algorithms and provide guidelines for selecting suitable algorithms in various scenarios. In particular, we introduce a learning framework that leverages the widely-used VQCs to perform the QPT task and integrate our AL algorithms into the query step. We demonstrate our algorithms by reconstructing the unitary quantum processes resulting from random quantum circuits with up to seven qubits. Numerical results show that our AL algorithms achieve significantly improved reconstruction compared to the baseline method that select a subset of quantum states randomly. Moreover, the improvement increases with the size of the underlying quantum system. Most importantly, our work introduces a new perspective for advancing existing QPT methods via active learning.

For future work, since we emphasize that our AL framework is general and can in principle be applied in the query step of any QPT method, one possible direction is to generalize it to other QPT methods beyond the variational quantum circuits considered here, such as tensor network QPT \cite{torlai2023quantum}. Another exciting direction is to extend the concept of active learning to other costly tasks in quantum information processing, such as shadow tomography, quantum detection tomography (QDT), quantum sensing, and quantum control \cite{gebhart2023learning}.

\begin{acknowledgments}
This work was partially supported by the Innovation Program for Quantum Science and Technology (Grant No.\ 2021ZD0302901), and National Natural Science Foundation of China (Grant No.\ 62102388).
\end{acknowledgments}

\nocite{*}

\bibliography{apssamp}

\end{document}